\documentclass[12pt]{iopart}

\usepackage{iopams}
\usepackage{graphicx}
\usepackage{bm}
\usepackage{amssymb}
\usepackage{color}
\usepackage[breaklinks=true,colorlinks=true,linkcolor=blue,urlcolor=blue,citecolor=blue]{hyperref}
\usepackage{dcolumn}
\usepackage{epstopdf}
\usepackage[dvipsnames]{xcolor}
\usepackage[utf8]{inputenc}
\usepackage[T1]{fontenc}
\usepackage{indentfirst}

\newcommand{\text}[1]{\mbox{\scriptsize{#1}}}
\newcommand{\tfrac}[2]{{\textstyle\frac{#1}{#2}}}

\newcommand{\ket}[1]{|#1\rangle}
\newcommand{\bra}[1]{\langle #1|}

\newcommand{\Fig}{figure}
\newcommand{\Figs}{figures}

\begin{document}

\title{Nuclear DFT analysis of electromagnetic moments in odd near doubly magic nuclei}

\author{P.L. Sassarini$^a$, J. Dobaczewski$^{a,b}$, J. Bonnard$^{a,c}$, R.F. Garcia Ruiz$^d$}
\address{
$^a$Department of Physics, University of York, Heslington, York YO10 5DD, United Kingdom \\
$^b$Institute of Theoretical Physics, Faculty of Physics, University of Warsaw, ul. Pasteura 5, PL-02-093 Warsaw, Poland \\
$^c$Universit{\'e} de Lyon, Institut de Physique des 2 Infinis de Lyon, IN2P3-CNRS-UCBL, 4 rue Enrico Fermi, 69622 Villeurbanne, France \\
$^d$Massachusetts Institute of Technology, Cambridge, MA 02139, USA}

\date{\today}

\ead{jacek.dobaczewski@york.ac.uk}

\begin{abstract}
We use the nuclear density functional theory to determine nuclear electric
quadrupole and magnetic dipole moments in all one-particle and
one-hole neighbours of eight doubly magic nuclei. We align angular
momenta along the intrinsic axial-symmetry axis with broken
time-reversal symmetry, which allows us to explore fully the
self-consistent charge, spin, and current polarisation.
Spectroscopic moments are determined for symmetry-restored wave
functions and compared with available experimental data. We find that
the obtained polarisations do not call for using quadrupole- or
dipole-moment operators with effective charges or effective
$g$-factors.
\end{abstract}

\submitto{\JPG}

\maketitle

Nuclear electromagnetic moments provide essential information in our
understanding of nuclear structure. Observables such as electric
quadrupole moments are highly sensitive to collective nuclear
phenomena~\cite{(Ney03),Bab16}, while magnetic dipole moments offer
sensitive probes to test our description of microscopic properties
such as those of valence nucleons~\cite{Pap13,(Gar15),McC19,Bou20,Dau21,(Ver22b)}
and the role of nuclear electroweak currents~\cite{Pas13,Car15}.
Although great progress was achieved in the description of
electromagnetic properties of light nuclei~\cite{Car15} and experimental trends in
certain isotopic chains, a unified and consistent description of
nuclear electromagnetic properties remains an open challenge for
nuclear theory.

Traditionally, shell model calculations were
successful in describing experimental trends of electromagnetic
moments~\cite{(Cas90b),(Ney03),Pap13,Klo19,(Rod20b)}. However, these calculations
require the use of effective nuclear charges and effective
$g$-factors that are most often phenomenologically adjusted
to data and sometimes lead to marked disagreements in different
regions of the nuclear chart~\cite{(Gar15),Gro17,rad18,(Lec21)}.
The nuclear DFT calculations in even-even nuclei have demonstrated a
good overall description of nuclear charge radii, see,
e.g., Refs.~\cite{Gar16,Gor19,Gro20,(Sca21)} and electric
quadrupole moments (in deformed nuclei BE2($2^+_1\rightarrow0^+_1$)
transitions)~\cite{(Del10a),(Sab07)} across the nuclear chart. An
analogous global DFT description of the magnetic dipole
moments~\cite{(Hof88),(Fur89),(Li13a),(Bon15),(Bor17),(Li18),(Per21),(Bar21)}
has not been fully developed so far.

In this Letter, we employ the nuclear DFT to address the challenge of
describing the electric and magnetic moments globally, that is,
without adjusting interactions, coupling constants, valence spaces,
or effective charges/$g$-factors separately in different regions of
the nuclear chart. In particular, our work provides the first global
adjustment of nuclear DFT's time-odd mean-field sector to magnetic dipole moments.
To minimize the possible pairing or beyond-mean-field effects,
such as, e.g., the coupling to collective motion, we chose to look at the
simplest possible cases of one-particle and one-hole neighbours of the
doubly magic nuclei. In addition, to check if the data indicate a
need for corrective terms, we used the simplest one-body
magnetic-moment operator. Here we note that the two-body-current
extensions of the Gamow-Teller transition operator were
shown to bring important improvement to the agreement with data
\cite{Pas13,(Gys19)}. Establishing the baseline for including analogous
corrections to the magnetic-moment operator is, therefore, of
paramount importance.
The deformed DFT approach is probably close to a spherical
Quasiparticle-Phonon Interaction
method~\cite{(Bor08a),(Bor10),(Ach14),(Co15),(Sap17),(Kam19),(Tse22)},
with an explicit coupling of spherical quasiparticles and phonons
replaced by the use of symmetry-restored deformed DFT configurations.
However, a dedicated comparative analysis of prospective links is
not available yet.

In nuclear DFT, properties of odd nuclei can be analyzed in terms of
the self-consistent polarisation effects caused by the presence of
the unpaired nucleon. Indeed, the non-zero quadrupole moment of the odd
nucleon induces deformation of the total mean field and thus
generates quadrupole moments of all remaining core nucleons. The latter
enhance the deformation of the mean field even more, which in
turn influences the quadrupole moment of the odd nucleon. In the
self-consistent solution, these mutual polarisations are effectively
summed up to infinity, whereupon the final total electric quadrupole moment $Q$
is generated~\cite{(Boh69a)}.

In a similar way, non-zero spin and current distributions of
the odd particle influence those of all other nucleons; in the
self-consistent solution they lead to a specific polarisation of the
system and give the total magnetic dipole moment $\mu$. All nucleons
contribute to the total moments, $Q$ and $\mu$, of the system, with their
individual contributions depending on polarisation responses to the
deformed and polarised mean fields. The broken-symmetry nuclear DFT
used here opens up a possibility of studying polarisation effects in
full. We note that approaches that start from unpolarised spherical
states are bound to treat the polarisation effects perturbatively, see,
e.g., Refs.~\cite{(Cas90b),(Co15),(Li18)}. In this Letter, we show that the
inclusion of self-consistent DFT spin and current polarisation
effects removes the necessity of introducing an effective spin
$g$-factor from the description of nuclear magnetic dipole moments.

In this work, we determined the electric quadrupole moments $Q$ and
magnetic dipole moments $\mu$ of 32 nuclei that are one-particle or
one-hole neighbours of eight doubly magic nuclei: $^{16}$O, $^{40}$Ca,
$^{48}$Ca, $^{56}$Ni, $^{78}$Ni, $^{100}$Sn, $^{132}$Sn, and
$^{208}$Pb. We employed code {\sc hfodd} (v3.07h)~\cite{(Dob21e)} for
three Skyrme functionals, UNEDF1~\cite{(Kor12b)},
SLy4~\cite{(Cha98a)}, and SkO$^\prime$~\cite{(Rei99a)}; for the Gogny
functional D1S~\cite{(Ber91e)}; and for the regularized functional
N$^3$LO (REG6d.190617)~\cite{(Ben20c)}.
We used the spherical basis of
$N_0=16$ shells.

We begun our analysis by selecting in doubly magic nuclei the
spherical orbitals that were closest to the Fermi energies, that is,
the lowest particle or highest hole orbitals were selected in
odd-particle or odd-hole nuclei, respectively. As it turned out, this
rule gave the same orbitals irrespective of which of the five
functionals were considered. In addition, this rule gave orbitals
corresponding to the ground-state spins and parities of odd nuclei,
whenever they were experimentally known. The exception was the
$I=\frac{3}{2}^+$ experimental ground state of $^{131}$Sn, whereas
the $I=\frac{11}{2}^-$ state was used in the present study.

Next, relative to the doubly magic nuclei, configurations of
odd-particle (odd-hole) nuclei were fixed by occupying (emptying)
deformed substates that originated from a given spherical orbital and
had the highest-positive (lowest-negative) value of $\Omega$, where
$\Omega$ denotes the eigenvalue of the projection of the
single-particle angular-momentum operator on the axial-symmetry axis.
The chosen single-particle occupations thus always corresponded to
the maximally aligned total angular momenta, $\Omega=+I$. Such
occupations yielded the oblate (prolate) self-consistent intrinsic
shapes for odd-particle (odd-hole) $I>\frac{1}{2}$ nuclei and vice
versa for $I=\frac{1}{2}$. For the configurations defined in this
way, fully self-consistent~\cite{(Rin80b),(Sch19)} unpaired solutions
were obtained for all odd nuclei considered in this work. The
convergence was stabilized by fixing specific partitions of
occupations among different $\Omega$ blocks, as described in detail
in~\cite{(Dob21e)}.

By occupying specific odd orbitals having good quantum numbers
$\Omega$, we aligned the angular momenta of odd unpaired nuclei along
the axial-symmetry axis of a deformed nucleus. This choice of
occupations requires explicit breaking of the time-reversal symmetry
in the intrinsic reference frame. As the direction of alignment with respect
to the nuclear shape strongly impacts the spin and current
polarisation effects~\cite{(Sat12a)}, this choice of occupations
constitutes an essential element of the DFT approach to magnetic
moments~\cite{(Hof88),(Fur89),(Bon15)}.

The aligned
configurations were strictly axial, which allowed us to determine the
angular-momentum-projected states $\ket{I M}$~\cite{(She21)} by
employing a one-dimensional integration over the Euler angle $\beta$
only,
\begin{equation}
\ket{I M} = N_{IM} \int_0^\pi  \mathrm{d}\beta\, d^I_{M\Omega} (\beta)
\exp\big( -\mathrm{i} \beta \hat{I}_y \bigr) \ket{\Omega},
\end{equation}
where $d^I_{M\Omega}$ are the Wigner functions~\cite{(Var88)},
$M$ is the projection of the
angular momentum on the laboratory $z$-axis, $\hat{I}_y$ denotes the $y$-component of the
angular-momentum operator, $\ket{\Omega}$ denotes the self-consistent deformed intrinsic state
defined above, and $N_{IM}$ is a normalization factor.
Instead of relying on approximate relations
between the intrinsic and spectroscopic nuclear moments, we
determined the standard magnetic $\mu$ and quadrupole $Q$ spectroscopic moments of the
angular-momentum-projected states $\ket{I M}$ as~\cite{(Rin80b)}
\begin{equation}
\label{spectroscopic}
\mu = \sqrt{\frac{4\pi}{3}}  \bra{II}\hat{M}_{10}\ket{II} , \quad
Q   = \sqrt{\frac{16\pi}{5}} \bra{II}\hat{Q}_{20}\ket{II} ,
\end{equation}
where $\hat{M}_{1\nu}$ and $\hat{Q}_{2\nu}$ are the $\nu$th magnetic components of
the corresponding M1 and E2 electromagnetic operators~\cite{(Dob00e)}, respectively.
In this way,  we could compare the calculated magnetic and quadrupole moments
directly with experimental data.

For the Skyrme functionals, one can separately adjust coupling
constants in the time-odd mean-field sector of the functional~\cite{(Per04c)}.
The impact of the time-odd mean-field sector of the Skyrme functional
on time-even observables (masses, radii, etc.), or single-particle
energies or spins and parities of ground states is small, and smaller
than overall deviations of these observables from data, see, e.g.,~\cite{(Pot10a)}.
Here we focus on studying the simplest terms in the time-odd mean-field sector, which
correspond to the spin-spin interactions
$\bm{\sigma}_1\cdot\bm{\sigma}_2$. Following~\cite{(Ben02d)}, we
parametrize these terms by the standard isoscalar and isovector
Landau parameters $g_0$ and $g_0'$, respectively. Since variations of
the isoscalar Landau parameter $g_0$ do not change the results, we
fixed it at $g_0=0.4$, which was the value recommended in~\cite{(Ben02d)}.

An important aspect of this work was to consistently follow given
configurations in function of the Landau parameter $g_0'$.
This was possible in all cases but for a few difficult ones
encountered for functional SkO$^\prime$. In particular, in
$^{101}$Sn, the converged solutions could not be obtained at any
value of $g_0'$ and in $^{47}$K at $g_0'>1.7$. This was caused by a
strong mixing with the neighboring deformed substates that had the
same values of $\Omega$. In addition, converged solutions obtained in
$^{57}$Ni and $^{57}$Cu turned out to have unusually large
deformations and thus became incomparable to those obtained for the
remaining four functionals studied here.

Theoretical uncertainties of the calculated magnetic moments were
estimated using the Bayesian Model Averaging (BMA) analysis
\cite{(Hoe99)}. In our implementation of the BMA, the unique variable
parameter was assumed to be the Landau parameter $g_0'$, for which,
based on Refs.~\cite{(Bor84),(Ben02d)}, we adopted a Gaussian prior
distribution of mean 1.2 and of variance 0.5. Theoretical BMA error bars
shown below correspond to the posterior estimates of the
uncertainties of the $g_0'$ optimum values obtained for Skyrme
functionals UNEDF1 and SLy4 in $^{101}$Sn, $^{47}$K, $^{57}$Ni, and
$^{57}$Cu and for Skyrme functionals UNEDF1, SLy4, and SkO$^\prime$
in all other nuclei.

Our study combines several aspects of the DFT approach to nuclear
moments that were never simultaneously considered so far. For
example, previous studies determined the intrinsic and not
spectroscopic moments~\cite{(Hof88),(Fur89),(Bon15),(Per21),(Bar21)},
used phenomenological description of the core
contributions~\cite{(Li13a),(Li18),(Per21),(Bar21)}, or neglected
time-reversal breaking~\cite{(Bor17),(Per21),(Bar21)} or
deformation~\cite{(Li13a),(Li18)} in the intrinsic reference frame.

\begin{figure}
\begin{center}
\includegraphics[width=0.6\columnwidth]{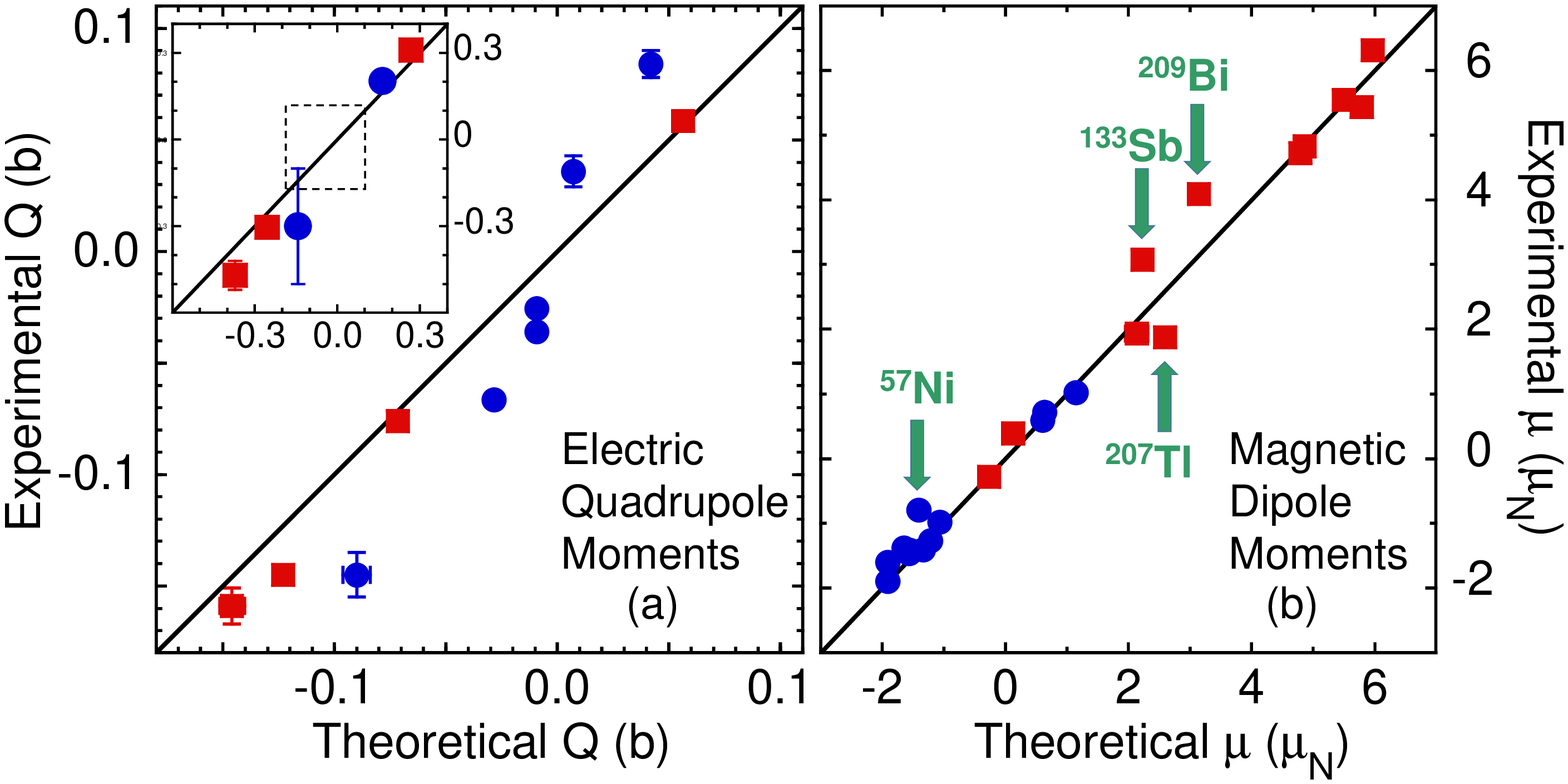}
\end{center}
\caption{Calculated electric quadrupole moments $Q$, panel (a),
compared with 15 experimentally measured values (the inset
shows values that are outside the area of the main plot, as
visualized by the dashed-line square drawn inside). Panel (b) shows
analogous results obtained for the magnetic dipole moments $\mu$
compared with 23 experimentally measured values (the arrows
mark the outlier cases discussed in the text). Full circles (squares)
show results obtained for $N$-odd ($Z$-odd) nuclei. Calculated values
shown in this figure were derived within the Bayesian Model Averaging
(BMA) analysis. Apart from one point, the
corresponding theoretical error bars are always smaller than the
sizes of symbols.
}
\label{fig:UNEDF1_Q}
\end{figure}

Before presenting details of our results, in {\Fig}~\ref{fig:UNEDF1_Q}
we show an overview of the comparison with the presently available
experimental data. We see that owing to the use of the whole
unabridged single-particle phase space, nuclear functionals properly
account for the charge and spin polarisations without invoking
effective charges or effective $g$-factors.

\newcommand{\notapp}{\multicolumn{1}{c} {n.a.}}
\newcommand{\notapr}{\multicolumn{1}{c|}{n.a.}}
\newcommand{\phs}{\phantom{$^*$}}
\newcommand{\phm}{\phantom{$-$}}
\renewcommand{\arraystretch}{1.4}
\begin{table*}
{\tiny\hspace*{+2cm}
\begin{tabular}{|r@{\,}r@{\,}l@{\,}r|l@{\,}r@{\,}r@{\,}r@{\,}r@{\,}r@{\,}r@{\,}l|}
\cline{5-12}
\multicolumn{4}{c|}{}             & \multicolumn{8}{c|}{Electric quadrupole moment $Q$ (b)}       \\
\hline
 Nuclide    & $I^\pi$    & $[Nn_z\Lambda]K$ & orbital    & ~~EXP~~~         & Ref.              & UNEDF1             &  SLy4    &SkO$^\prime$&   D1S    & N$^3$LO   & ~~Average      \\ 
\hline                                                                                                                                                                                      
 $^{17 }$O  & $\tfrac{5 }{2}^{+}$ &[202]5/2 & 1$d_{5/2}$ & $-$0.0256(2)$^*$ &\cite{(Sto16)} &          $-$0.0108 & $-$0.0087& $-$0.0086 & $-$0.0085 & $-$0.0098 & $-$0.0093(9)       \\ 
 $^{17 }$F  & $\tfrac{5 }{2}^{+}$ &[202]5/2 & 1$d_{5/2}$ & $-$0.076(4)$^*$  &\cite{(Sto16)} &          $-$0.0712 & $-$0.0721& $-$0.0720 & $-$0.0730 & $-$0.0691 & $-$0.0715(13)      \\ 
 $^{39 }$Ca & $\tfrac{3 }{2}^{+}$ &[202]3/2 & 1$d_{3/2}$ &\phm0.036(7)      &\cite{(Sto16)} &             0.0075 &    0.0070&    0.0070 &    0.0073 &    0.0072 &\phm0.0072(2)       \\ 
 $^{41 }$Ca & $\tfrac{7 }{2}^{-}$ &[303]7/2 & 1$f_{7/2}$ & $-$0.0665(18)    &\cite{(Sto16)} &          $-$0.0323 & $-$0.0270& $-$0.0270 & $-$0.0263 & $-$0.0288 & $-$0.028(2)        \\ 
 $^{39 }$K  & $\tfrac{3 }{2}^{+}$ &[202]3/2 & 1$d_{3/2}$ &\phm0.0585(6)     &\cite{(Sto16)} &             0.0546 &    0.0580&    0.0565 &    0.0576 &    0.0555 &\phm0.0564(13)      \\ 
 $^{41 }$Sc & $\tfrac{7 }{2}^{-}$ &[303]7/2 & 1$f_{7/2}$ & $-$0.145(3)$^*$  &\cite{(Sto16)} &          $-$0.1199 & $-$0.1255& $-$0.1218 & $-$0.1266 & $-$0.1211 & $-$0.123(3)        \\ 
 $^{47 }$Ca & $\tfrac{7 }{2}^{-}$ &[303]7/2 & 1$f_{7/2}$ &\phm0.084(6)      &\cite{(Gar15)} &             0.0460 &    0.0395&    0.0441 &    0.0379 &    0.0415 &\phm0.042(3)        \\ 
 $^{49 }$Ca & $\tfrac{3 }{2}^{-}$ &[301]3/2 & 2$p_{3/2}$ & $-$0.036(3)      &\cite{(Gar15)} &          $-$0.0104 & $-$0.0084& $-$0.0094 & $-$0.0079 & $-$0.0103 & $-$0.0093(10)      \\ 
 $^{49 }$Sc & $\tfrac{7 }{2}^{-}$ &[303]7/2 & 1$f_{7/2}$ & $-$0.159(8)      &\cite{(Bai22)} &          $-$0.1545 & $-$0.1455& $-$0.1496 & $-$0.1408 & $-$0.1393 & $-$0.146(6)        \\ 
 $^{55 }$Ni & $\tfrac{7 }{2}^{-}$ &[303]7/2 & 1$f_{7/2}$ &                  &               &             0.1637 &    0.1486&    0.1661 &    0.1274 &    0.1336 &\phm0.148(16)       \\ 
 $^{57 }$Ni & $\tfrac{3 }{2}^{-}$ &[301]3/2 & 2$p_{3/2}$ &                  &               &          $-$0.0685 & $-$0.0511& $-$0.1622 & $-$0.0434 & $-$0.0518 & $-$0.054(9)$^{**}$ \\ 
 $^{55 }$Co & $\tfrac{7 }{2}^{-}$ &[303]7/2 & 1$f_{7/2}$ &                  &               &             0.2254 &    0.2241&    0.2371 &    0.2086 &    0.2091 &\phm0.221(11)       \\ 
 $^{57 }$Cu & $\tfrac{3 }{2}^{-}$ &[301]3/2 & 2$p_{3/2}$ &                  &               &          $-$0.1207 & $-$0.1143& $-$0.1919 & $-$0.1077 & $-$0.1096 & $-$0.113(5)$^{**}$ \\ 
 $^{77 }$Ni & $\tfrac{9 }{2}^{+}$ &[404]9/2 & 1$g_{9/2}$ &                  &               &             0.1600 &    0.1305&    0.1555 &    0.1197 &    0.1275 &\phm0.139(16)       \\ 
 $^{79 }$Ni & $\tfrac{5 }{2}^{+}$ &[402]5/2 & 2$d_{5/2}$ &                  &               &          $-$0.0797 & $-$0.0601& $-$0.0825 & $-$0.0513 & $-$0.0581 & $-$0.066(13)       \\ 
 $^{77 }$Co & $\tfrac{7 }{2}^{-}$ &[303]7/2 & 1$f_{7/2}$ &                  &               &             0.2075 &    0.1847&    0.2121 &    0.1874 &    0.1778 &\phm0.194(13)       \\ 
 $^{79 }$Cu & $\tfrac{3 }{2}^{-}$ &[301]3/2 & 2$p_{3/2}$ &                  &               &          $-$0.1033 & $-$0.0962& $-$0.0853 & $-$0.0940 & $-$0.0910 & $-$0.094(6)        \\ 
 $^{99 }$Sn & $\tfrac{9 }{2}^{+}$ &[404]9/2 & 1$g_{9/2}$ &                  &               &             0.1719 &    0.1628&    0.1773 &    0.1507 &    0.1575 &\phm0.164(10)       \\ 
 $^{101}$Sn & $\tfrac{5 }{2}^{+}$ &[402]5/2 & 2$d_{5/2}$ &                  &               &          $-$0.0927 & $-$0.0842&           & $-$0.0788 & $-$0.0920 & $-$0.0870(6)$^{**}$\\ 
 $^{99 }$In & $\tfrac{9 }{2}^{+}$ &[404]9/2 & 1$g_{9/2}$ &                  &               &             0.2848 &    0.2935&    0.3040 &    0.2865 &    0.2848 &\phm0.291(7)        \\ 
 $^{101}$Sb & $\tfrac{7 }{2}^{+}$ &[404]7/2 & 1$g_{7/2}$ &                  &               &          $-$0.2936 & $-$0.2975& $-$0.2858 & $-$0.2921 & $-$0.2903 & $-$0.292(4)        \\ 
 $^{131}$Sn & $\tfrac{11}{2}^{-}$ &[505]11/2& 1$h_{11/2}$&\phm0.203(4)      &\cite{(Yor20)} &             0.1737 &    0.1616&    0.1780 &    0.1507 &    0.1596 &\phm0.165(10)       \\ 
 $^{133}$Sn & $\tfrac{7 }{2}^{-}$ &[503]7/2 & 2$f_{7/2}$ & $-$0.145(10)     &\cite{(Rod20b)}&          $-$0.0919 & $-$0.0845& $-$0.0979 & $-$0.0815 & $-$0.0941 & $-$0.090(6)        \\ 
 $^{131}$In & $\tfrac{9 }{2}^{+}$ &[404]9/2 & 1$g_{9/2}$ &\phm0.31(1)       &\cite{(Ver22b)}&             0.2615 &    0.2664&    0.2815 &    0.2712 &    0.2589 &\phm0.268(8)        \\ 
 $^{133}$Sb & $\tfrac{7 }{2}^{+}$ &[404]7/2 & 1$g_{7/2}$ & $-$0.304(7)      &\cite{(Lec21)} &          $-$0.2549 & $-$0.2566& $-$0.2503 & $-$0.2609 & $-$0.2508 & $-$0.255(4)        \\ 
 $^{209}$Pb & $\tfrac{9 }{2}^{+}$ &[604]9/2 & 2$g_{9/2}$ & $-$0.27(17)      &\cite{(Sto16)} &          $-$0.1514 & $-$0.1450& $-$0.1348 & $-$0.1325 & $-$0.1510 & $-$0.143(8)        \\ 
 $^{209}$Bi & $\tfrac{9 }{2}^{-}$ &[505]9/2 & 1$h_{9/2}$ & $-$0.47(5)$^{***}$&\cite{(Sto16),(Skr21)}& $-$0.3710 & $-$0.3736& $-$0.3661 & $-$0.3835 & $-$0.3643 & $-$0.372(7)         \\ 
\hline
\multicolumn{12}{l}{$^*$sign not measured; calculated sign was assigned.} \\
\multicolumn{12}{l}{$^{**}$functional SkO$^\prime$ excluded.}             \\
\multicolumn{12}{l}{$^{***}$average of $-$0.516(15) \cite{(Sto16)} and $-$0.418(6) \cite{(Skr21)}
with the error bar reflecting uncertainties of the atomic} \\
\multicolumn{12}{l}{~~~~theory.}
\end{tabular}
}
\caption{Experimental values of the electric quadrupole
moments $Q$ compared with those calculated for functionals UNEDF1, SLy4, SkO$^\prime$, D1S, and N$^3$LO.
}
\label{tab:UNEDF1_Q}
\end{table*}

\begin{table*}
{\tiny\hspace*{+1.3cm}
\begin{tabular}{|r@{\,}r@{\,}l@{\,}r|l@{\,}r@{\,}r@{\,}r@{\,}r@{\,}r@{\,}r@{\,}l@{\,}l|}
\cline{5-13}
\multicolumn{4}{c|}{}             & \multicolumn{9}{c|}{Magnetic dipole moment $\mu$ ($\mu_N$)}  \\
\hline
 Nuclide    & $I^\pi$    & $[Nn_z\Lambda]K$ & orbital    & ~~EXP               & Ref.                & UNEDF1    &  SLy4     &SkO$^\prime$&   D1S    & N$^3$LO   & ~~Average        & ~~BMA             \\ 
\hline                                                                                                                                                                                                     
 $^{15 }$O  & $\tfrac{1 }{2}^{-}$ &[101]1/2 & 1$p_{1/2}$ &\phm0.71951(12)$^*$  & \cite{(Sto05d)} &    0.6366 &    0.6372 &    0.6384 &    0.6369 &    0.6352 &\phm0.6369(10)    &\phm0.6375(8)      \\ 
 $^{17 }$O  & $\tfrac{5 }{2}^{+}$ &[202]5/2 & 1$d_{5/2}$ & $-$1.89379(9)$^*$   & \cite{(Sto05d)} & $-$1.9081 & $-$1.9092 & $-$1.9090 & $-$1.9098 & $-$1.9091 & $-$1.9090(6)     & $-$1.9087(5)      \\ 
 $^{15 }$N  & $\tfrac{1 }{2}^{-}$ &[101]1/2 & 1$p_{1/2}$ & $-$0.2830569(14)$^*$& \cite{(Ant05a)} & $-$0.2632 & $-$0.2638 & $-$0.2651 & $-$0.2632 & $-$0.2616 & $-$0.2634(12)    & $-$0.2642(8)      \\ 
 $^{17 }$F  & $\tfrac{5 }{2}^{+}$ &[202]5/2 & 1$d_{5/2}$ &\phm4.7223(12)$^*$   & \cite{(Sto05d)} &    4.7878 &    4.7890 &    4.7881 &    4.7895 &    4.7889 &\phm4.7887(6)     &\phm4.7882(6)      \\ 
 $^{39 }$Ca & $\tfrac{3 }{2}^{+}$ &[202]3/2 & 1$d_{3/2}$ &\phm1.02168(12)      & \cite{(Sto05d)} &    1.1465 &    1.1468 &    1.1476 &    1.1472 &    1.1469 &\phm1.1470(4)     &\phm1.1470(5)      \\ 
 $^{41 }$Ca & $\tfrac{7 }{2}^{-}$ &[303]7/2 & 1$f_{7/2}$ & $-$1.5942(7)        & \cite{(Sto05d)} & $-$1.9088 & $-$1.9099 & $-$1.9098 & $-$1.9098 & $-$1.9098 & $-$1.9096(4)     & $-$1.9096(4)      \\ 
 $^{39 }$K  & $\tfrac{3 }{2}^{+}$ &[202]3/2 & 1$d_{3/2}$ &\phm0.39147(3)       & \cite{(Sto05d)} &    0.1259 &    0.1256 &    0.1249 &    0.1251 &    0.1255 &\phm0.1254(4)     &\phm0.1254(4)      \\ 
 $^{41 }$Sc & $\tfrac{7 }{2}^{-}$ &[303]7/2 & 1$f_{7/2}$ &\phm5.431(2)$^*$     & \cite{(Sto05d)} &    5.7886 &    5.7897 &    5.7888 &    5.7896 &    5.7895 &\phm5.7892(5)     &\phm5.7890(5)      \\ 
 $^{47 }$Ca & $\tfrac{7 }{2}^{-}$ &[303]7/2 & 1$f_{7/2}$ & $-$1.4064(11)       & \cite{(Gar15)}  & $-$1.4113 & $-$1.3232 & $-$1.2991 & $-$1.4894 & $-$1.4248 & $-$1.39(7)       & $-$1.33(10)       \\ 
 $^{49 }$Ca & $\tfrac{3 }{2}^{-}$ &[301]3/2 & 2$p_{3/2}$ & $-$1.3799(8)        & \cite{(Gar15)}  & $-$1.6506 & $-$1.6494 & $-$1.6530 & $-$1.7090 & $-$1.6607 & $-$1.66(2)       & $-$1.65(4)        \\ 
 $^{47 }$K  & $\tfrac{1 }{2}^{+}$ &[220]1/2 & 2$s_{1/2}$ &\phm1.933(9)         & \cite{(Sto05d)} &    2.2040 &    2.0786 &    2.2958 &    2.4801 &    2.5769 &\phm2.33(18)      &\phm2.13(7)$^{**}$ \\ 
 $^{49 }$Sc & $\tfrac{7 }{2}^{-}$ &[303]7/2 & 1$f_{7/2}$ &\phm5.539(4)         & \cite{(Bai22)}  &    5.3409 &    5.4636 &    5.6621 &    5.6127 &    5.4734 &\phm5.51(11)      &\phm5.50(14)       \\ 
 $^{55 }$Ni & $\tfrac{7 }{2}^{-}$ &[303]7/2 & 1$f_{7/2}$ & $-$0.98(3)$^*$      & \cite{(Ber09b)} & $-$1.1009 & $-$1.0923 & $-$1.0309 & $-$1.3596 & $-$1.1638 & $-$1.15(11)      & $-$1.06(13)       \\ 
 $^{57 }$Ni & $\tfrac{3 }{2}^{-}$ &[301]3/2 & 2$p_{3/2}$ & $-$0.7975(14)$^*$   & \cite{(Sto05d)} & $-$1.3267 & $-$1.4371 & $-$0.4178 & $-$1.5227 & $-$1.4261 & $-$1.43(7)$^{**}$& $-$1.40(8)$^{**}$ \\ 
 $^{55 }$Co & $\tfrac{7 }{2}^{-}$ &[303]7/2 & 1$f_{7/2}$ &\phm4.822(3)$^*$     & \cite{(Sto05d)} &    4.9296 &    4.8991 &    4.8016 &    5.1811 &    4.9969 &\phm4.96(13)      &\phm4.86(15)       \\ 
 $^{57 }$Cu & $\tfrac{3 }{2}^{-}$ &[301]3/2 & 2$p_{3/2}$ &\phm                 &                 &    3.1759 &    3.2968 &    2.0319 &    3.4081 &    3.2944 &\phm3.29(8)$^{**}$&\phm3.26(8)$^{**}$ \\ 
 $^{77 }$Ni & $\tfrac{9 }{2}^{+}$ &[404]9/2 & 1$g_{9/2}$ &\phm                 &                 & $-$1.2069 & $-$1.1768 & $-$1.1414 & $-$1.4322 & $-$1.2563 & $-$1.24(10)      & $-$1.16(13)       \\ 
 $^{79 }$Ni & $\tfrac{5 }{2}^{+}$ &[402]5/2 & 2$d_{5/2}$ &\phm                 &                 & $-$1.5128 & $-$1.5542 & $-$1.4924 & $-$1.6529 & $-$1.5754 & $-$1.56(6)       & $-$1.51(8)        \\ 
 $^{77 }$Co & $\tfrac{7 }{2}^{-}$ &[303]7/2 & 1$f_{7/2}$ &\phm                 &                 &    4.9185 &    4.9234 &    4.7569 &    5.1730 &    4.9936 &\phm4.95(13)      &\phm4.85(16)       \\ 
 $^{79 }$Cu & $\tfrac{3 }{2}^{-}$ &[301]3/2 & 2$p_{3/2}$ &\phm                 &                 &    3.2102 &    3.3391 &    3.3742 &    3.4565 &    3.3927 &\phm3.35(8)       &\phm3.31(8)        \\ 
 $^{99 }$Sn & $\tfrac{9 }{2}^{+}$ &[404]9/2 & 1$g_{9/2}$ &\phm                 &                 & $-$1.2018 & $-$1.1918 & $-$1.1477 & $-$1.4448 & $-$1.2608 & $-$1.25(10)      & $-$1.17(13)       \\ 
 $^{101}$Sn & $\tfrac{5 }{2}^{+}$ &[402]5/2 & 2$d_{5/2}$ &\phm                 &                 & $-$1.4674 & $-$1.4968 &           & $-$1.5824 & $-$1.4956 & $-$1.51(4)$^{**}$& $-$1.50(8)$^{**}$ \\ 
 $^{99 }$In & $\tfrac{9 }{2}^{+}$ &[404]9/2 & 1$g_{9/2}$ &\phm                 &                 &    6.0398 &    6.0097 &    5.9342 &    6.2765 &    6.1021 &\phm6.07(12)      &\phm5.98(14)       \\ 
 $^{101}$Sb & $\tfrac{7 }{2}^{+}$ &[404]7/2 & 1$g_{7/2}$ &\phm                 &                 &    2.2721 &    2.1716 &    2.1604 &    2.1313 &    2.1465 &\phm2.18(5)       &\phm2.20(8)        \\ 
 $^{131}$Sn & $\tfrac{11}{2}^{-}$ &[505]11/2& 1$h_{11/2}$& $-$1.267(1)         & \cite{(Yor20)}  & $-$1.2443 & $-$1.2301 & $-$1.2174 & $-$1.4868 & $-$1.3184 & $-$1.30(10)      & $-$1.22(13)       \\ 
 $^{133}$Sn & $\tfrac{7 }{2}^{-}$ &[503]7/2 & 2$f_{7/2}$ & $-$1.410(1)         & \cite{(Rod20b)} & $-$1.5391 & $-$1.5607 & $-$1.5775 & $-$1.6580 & $-$1.5713 & $-$1.58(4)       & $-$1.55(8)        \\ 
 $^{131}$In & $\tfrac{9 }{2}^{+}$ &[404]9/2 & 1$g_{9/2}$ &\phm6.312(14)        & \cite{(Ver22b)} &    6.0340 &    6.0133 &    5.9055 &    6.2650 &    6.0926 &\phm6.06(12)      &\phm5.97(15)       \\ 
 $^{133}$Sb & $\tfrac{7 }{2}^{+}$ &[404]7/2 & 1$g_{7/2}$ &\phm3.070(2)         & \cite{(Lec21)}  &    2.2813 &    2.1792 &    2.2088 &    2.1125 &    2.1379 &\phm2.18(6)       &\phm2.23(9)        \\ 
 $^{207}$Pb & $\tfrac{1 }{2}^{-}$ &[501]1/2 & 3$p_{1/2}$ &\phm0.5906(4)        & \cite{(Adr16)}  &    0.6059 &    0.6021 &    0.6129 &    0.6120 &    0.5972 &\phm0.606(6)      &\phm0.606(11)      \\ 
 $^{209}$Pb & $\tfrac{9 }{2}^{+}$ &[604]9/2 & 2$g_{9/2}$ & $-$1.4735(16)       & \cite{(Sto05d)} & $-$1.5270 & $-$1.5539 & $-$1.6222 & $-$1.6556 & $-$1.5664 & $-$1.59(5)       & $-$1.56(9)        \\ 
 $^{207}$Tl & $\tfrac{1 }{2}^{+}$ &[400]1/2 & 3$s_{1/2}$ &\phm1.876(5)         & \cite{(Sto05d)} &    2.5797 &    2.6036 &    2.6051 &    2.6475 &    2.6135 &\phm2.61(2)       &\phm2.59(4)        \\ 
 $^{209}$Bi & $\tfrac{9 }{2}^{-}$ &[505]9/2 & 1$h_{9/2}$ &\phm4.092(2)         & \cite{(Skr18)}  &    3.2065 &    3.1027 &    3.1249 &    3.0136 &    3.0522 &\phm3.10(7)       &\phm3.15(9)        \\ 
\hline
\multicolumn{13}{l}{$^*$sign not measured; calculated sign was assigned.} \\
\multicolumn{13}{l}{$^{**}$functional SkO$^\prime$ excluded.}
\end{tabular}
}
\caption{Same as in Table~\protect\ref{tab:UNEDF1_Q} but for the magnetic dipole moments $\mu$.
}
\label{tab:UNEDF1_M}
\end{table*}

In Tables~\ref{tab:UNEDF1_Q} and~\ref{tab:UNEDF1_M}, we collected all
results obtained for electric quadrupole and magnetic dipole moments,
respectively. There we also show Nilsson labels and orbitals
corresponding to the configurations used, experimental data along
with the corresponding references to the original publications or
compilations thereof, and averages and RMS deviations of results
obtained for the five functionals used in this study. In
Table~\ref{tab:UNEDF1_M}, we also show results obtained within the
BMA analysis.

As indicated above, we used the standard single-particle magnetic-dipole-moment operator,
\begin{equation}
\hat{\bm{\mu}} \equiv \hat{\bm{M}}_1 = g_\ell^p   \hat{\bm{L}}_p   +g_s^n  \hat{\bm{S}}_n
                                            +g_s^p  \hat{\bm{S}}_p  ,
\end{equation}
where $\hat{\bm{L}}_p$ is the proton orbital angular-momentum operator and
$\hat{\bm{S}}_p$ ($\hat{\bm{S}}_n$)
is the proton (neutron) spin operator,
and where the
bare proton and neutron orbital and spin gyromagnetic factors read
\begin{equation}
\label{eq:miu}
g_\ell^p=\mu_N, ~g_s^n=-3.826\,\mu_N, ~g_s^p=+5.586\,\mu_N.
\end{equation}
Since the total angular momentum $\hat{\bm{I}}=\sum_{\nu=n,p}(\hat{\bm{L}}_\nu +\hat{\bm{S}}_\nu)$
is conserved,
it is convenient to subtract $I\,\mu_N$ from the spectroscopic magnetic dipole moments of
odd-$Z$ nuclei while leaving those of odd-$N$ nuclei unchanged.
In this way, we define the "spin" magnetic dipole moments as
\begin{eqnarray}
\label{eq:spin1}
\mu^{\text{S}}_{\text{odd-$Z$}} &\equiv& \langle\hat{\bm{\mu}}\rangle - I\,\mu_N
       = g_\ell^{n\prime} \langle\hat{\bm{L}}_n\rangle
           + g_s^   {n\prime} \langle\hat{\bm{S}}_n\rangle
           + g_s^   {p\prime} \langle\hat{\bm{S}}_p\rangle,  \\
\label{eq:spin2}
\mu^{\text{S}}_{\text{odd-$N$}} &\equiv&  \langle\hat{\bm{\mu}}\rangle \hspace*{11.5mm}
         = g_\ell^p \langle\hat{\bm{L}}_p\rangle
           + g_s^n    \langle\hat{\bm{S}}_n\rangle
           + g_s^p    \langle\hat{\bm{S}}_p\rangle,
\end{eqnarray}
where symbols $\langle\rangle$ denote the standard matrix elements that
define spectroscopic moments (\ref{spectroscopic}), and
\begin{equation}
g_\ell^{n\prime}=-\mu_N,~
   g_s^{n\prime}=-4.826\,\mu_N,~
   g_s^{p\prime}=+4.586\,\mu_N.
\end{equation}

\begin{figure}
\begin{center}
\includegraphics[width=0.6\columnwidth]{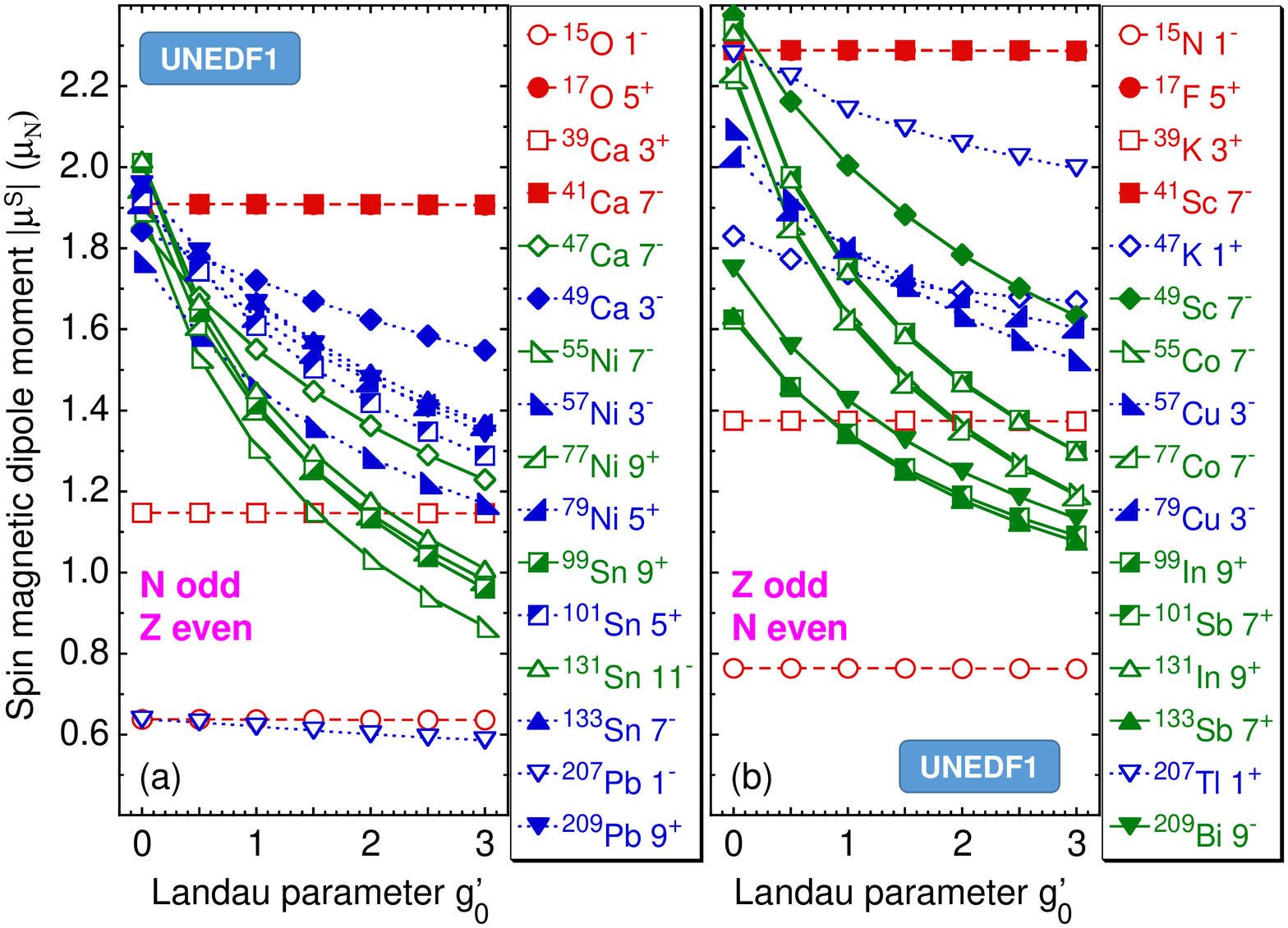}
\end{center}
\caption{Absolute values of the UNEDF1 spin magnetic dipole moments $|\mu^{\text{S}}$|,
Eqs.~(\protect\ref{eq:spin1})--(\protect\ref{eq:spin2}), calculated
in function of the Landau parameter $g_0'$ for $N$-odd-$Z$-even (a)
and $Z$-odd-$N$-even (b) nuclei. Dashed, solid, and dotted lines denote nuclei belonging to
the first, second, and third group discussed in the text, respectively.
Values obtained for $^{17}$O and
$^{17}$F are hidden behind those obtained for  $^{41}$Ca and
$^{41}$Sc, respectively. Doubled ground-state spin and parity are
given in the legends. Full and empty symbols denote particle and hole
states, respectively.
}
\label{fig:UNEDF1_Mg0_all}
\end{figure}

The spin magnetic dipole moments $\mu^{\text{S}}$
(\ref{eq:spin1})--(\ref{eq:spin2}) can be trivially evaluated for
experimental and theoretical results. They allow for comparing
odd-$N$ and odd-$Z$ nuclei on an equal footing and in the same scale.
Moreover, values of $\mu^{\text{S}}$ obey a simple rule that defines
their signs, namely, those for an odd proton in a
$j=\ell+\tfrac{1}{2}$ configuration or for an odd neutron in a
$j=\ell-\tfrac{1}{2}$ configuration are positive and otherwise, they
are negative. Therefore, without any loss of information, we can
meaningfully plot and compare the absolute values $|\mu^{\text{S}}|$
only. In this way, by plotting $|\mu^{\text{S}}|$ in {\Fig}~\ref{fig:UNEDF1_Mg0_all}
we can show and discuss our results in a much finer scale than that used in
the overview figure~\ref{fig:UNEDF1_Q}(b).

\begin{figure}
\begin{center}
\includegraphics[width=0.6\columnwidth]{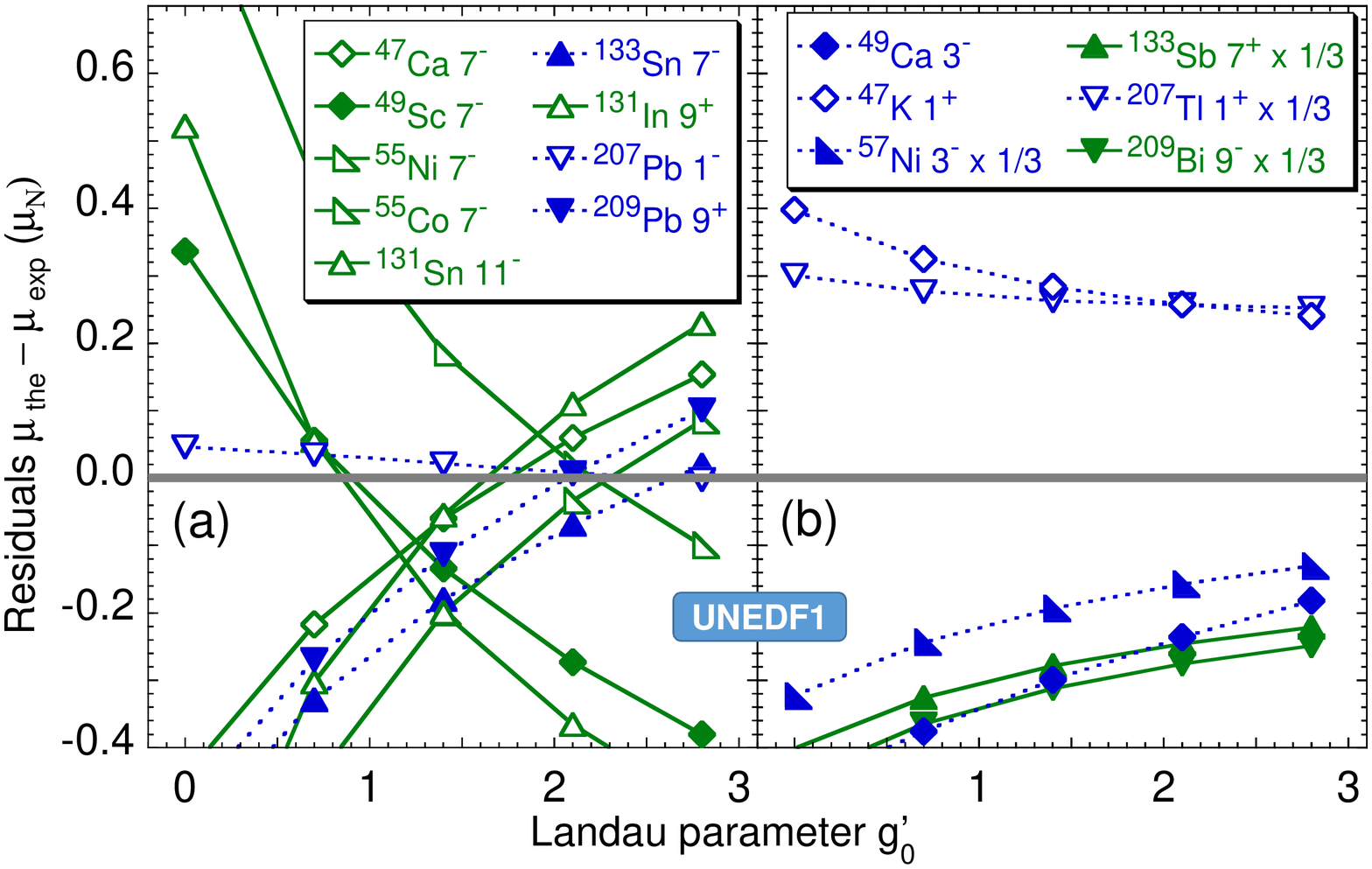}
\end{center}
\caption{The UNEDF1 magnetic dipole moments $\mu$ calculated in function of
the Landau parameter $g_0'$ relative to the experimental values.
Panels (a) and (b) show results that do and do not cross the line of
$\mu_{\text{the}}=\mu_{\text{exp}}$, respectively.
Solid and dotted lines denote nuclei belonging to
the second and third group discussed in the text, respectively.
Symbols $\times1/3$ denote outlier values multiplied by a factor of 1/3 to fit in
the scale of the figure.
}
\label{fig:UNEDF1_Mg0}
\end{figure}

The main thrust of our study is in
the dependence of the results on the isovector Landau parameter $g_0'$.
This is what we show in {\Figs}~\ref{fig:UNEDF1_Mg0_all}
and~\ref{fig:UNEDF1_Mg0} for the UNEDF1 spin magnetic dipole moments
$|\mu^{\text{S}}|$ and residuals $\mu_{\text{the}}-\mu_{\text{exp}}$,
respectively.
The magnetic dipole moments obtained for the 32 nuclei
considered in this study allow for grouping them into three distinct
sets that illustrate the mechanism of the spin polarisation induced by
the isovector spin-spin interaction.

The first group (dashed lines in {\Fig}~\ref{fig:UNEDF1_Mg0_all})
contains eight lightest nuclei around $^{16}$O and
$^{40}$Ca, which are characterized by all spin-orbit partners located
on the same side of the Fermi energy. In these nuclei, irrespective
of whether an odd proton or an odd neutron, or a hole or particle
state, or a high or low spin state are occupied, no tangible
polarisation of the spin distribution is obtained and no ensuing
dependence of the magnetic dipole moment on the isovector spin-spin interaction is
visible. As a result, in this group, all magnetic dipole moments stay quite
rigidly fixed at the Schmidt limits~\cite{(Sch37)}. Corrections to
the Schmidt values were already studied, see, e.g.,~\cite{(Wei12)},
with a moderate level of success in describing the experimental data.

The second group (solid lines in {\Figs}~\ref{fig:UNEDF1_Mg0_all}
and~\ref{fig:UNEDF1_Mg0}) contains nuclei around heavier doubly magic nuclei,
which are characterized by the Fermi energies separating pairs of the
spin-orbit partners from one another, and by a hole or a particle
created in one of the spin-orbit partners. In all such nuclei, irrespective of
whether the nucleus contains an odd-proton or an odd-neutron, the
dependence of the magnetic dipole moments on the isovector spin-spin interaction is strong. Two exceptions
from this rule are the cases of $^{47}$Ca and $^{49}$Sc, where
only the neutron pair of spin-orbit partners is available for polarisation
and the response to the isovector spin-spin interaction is somewhat weaker. With increasing
values of $g_0'$, the calculated magnetic dipole moments significantly
depart from the Schmidt limits.

The third group (dotted lines in {\Figs}~\ref{fig:UNEDF1_Mg0_all}
and~\ref{fig:UNEDF1_Mg0}) contains nuclei in which particles or holes are created
in non-intruder states or their partners. Then, the spin polarisation of
the spin-orbit partners becomes weaker and, as a result, the
dependence of the magnetic dipole moments on the isovector spin-spin interaction weakens too.
For $\tfrac{1}{2}^\pm$ states, such dependence is particularly weak.

\begin{figure}
\begin{center}
\includegraphics[width=0.6\columnwidth]{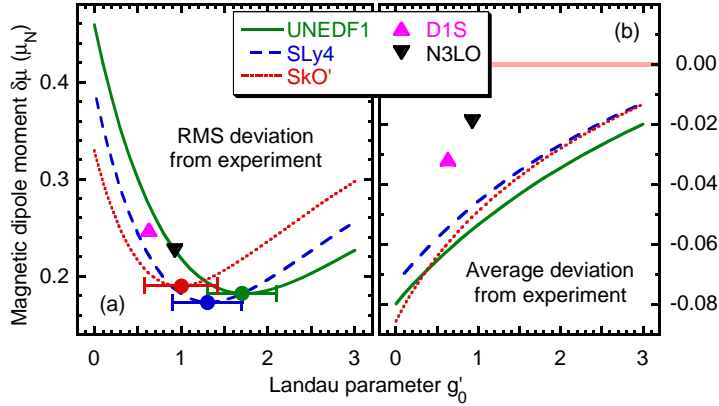}
\end{center}
\caption{RMS (a) and average (b) deviations $\delta\mu$ between
the calculated and experimental values of magnetic dipole moments.
}
\label{fig:UNEDF1_D}
\end{figure}

\begin{figure}
\begin{center}
\includegraphics[width=0.6\columnwidth]{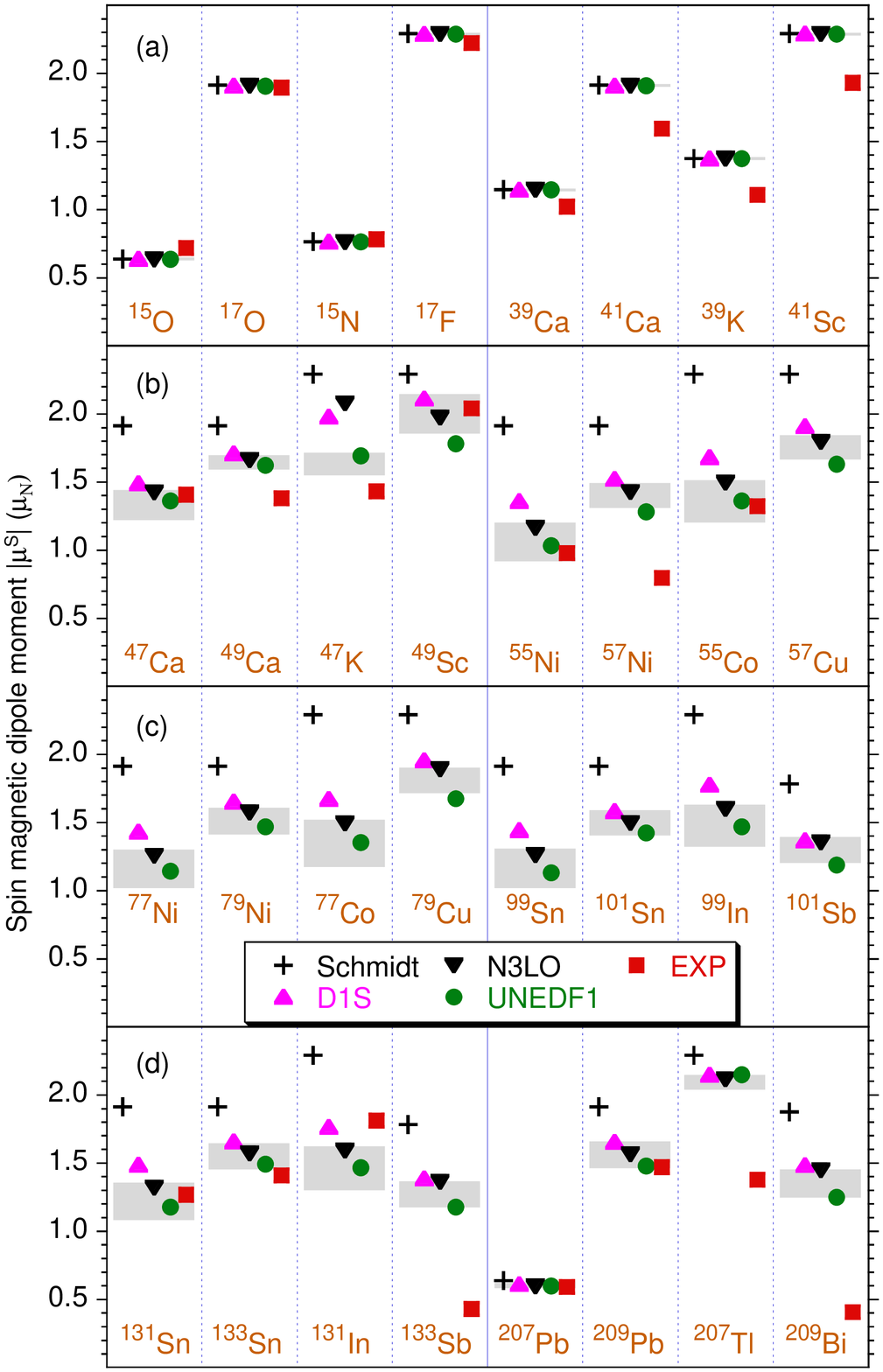}
\end{center}
\caption{Absolute values of the spin magnetic dipole moments $|\mu^{\text{S}}|$,
Eqs.~(\protect\ref{eq:spin1})--(\protect\ref{eq:spin2}), calculated
for the D1S, N$^3$LO, and UNEDF1 functionals and compared with the
Schmidt values and experimental data.
}
\label{fig:UNEDF1_M}
\end{figure}

In {\Figs}~\ref{fig:UNEDF1_Mg0}(a) and~\ref{fig:UNEDF1_Mg0}(b), we show nuclei whose
magnetic dipole moments as functions of $g_0'$ can and cannot, respectively, be brought
down to experimental data. In particular,
$^{57}$Ni, $^{133}$Sb, $^{207}$Tl, and $^{209}$Bi are
clear outliers, with the calculated magnetic dipole moments strongly
deviating from experiment, see discussion below.

For the other two Skyrme functionals considered in this work, SLy4
and SkO$^\prime$, the pattern of dependence of the magnetic dipole moments
on the Landau parameter $g_0'$ is fairly similar. The optimum values of
the Landau parameter $g_0'$, for which the RMS deviations between the
calculated and measured magnetic dipole moments are smallest,
see the full dots in {\Fig}~\ref{fig:UNEDF1_D}(a), are equal to $g_0'=1.0$, 1.3, and 1.7 for
SkO$^\prime$, SLy4, and UNEDF1, respectively. Below we present
results calculated at these particular values of $g_0'$. It is
rewarding to see that within the estimated uncertainties, the optimum
values are not only compatible with one another but also with the
estimate derived in Refs.~\cite{(Bor84),(Ben02d)} from the analysis
of the Gamow-Teller resonances and $\beta$ decays.

In {\Fig}~\ref{fig:UNEDF1_M} we show the complete set of our magnetic
dipole moments calculated for the D1S, N$^3$LO, and UNEDF1
functionals and compared with the Schmidt values and experimental
data. The shaded bands displayed in the
figure correspond to the averages and $1\sigma$ deviations computed within the
BMA for the three considered Skyrme functionals.

\begin{figure}
\begin{center}
\includegraphics[width=0.6\columnwidth]{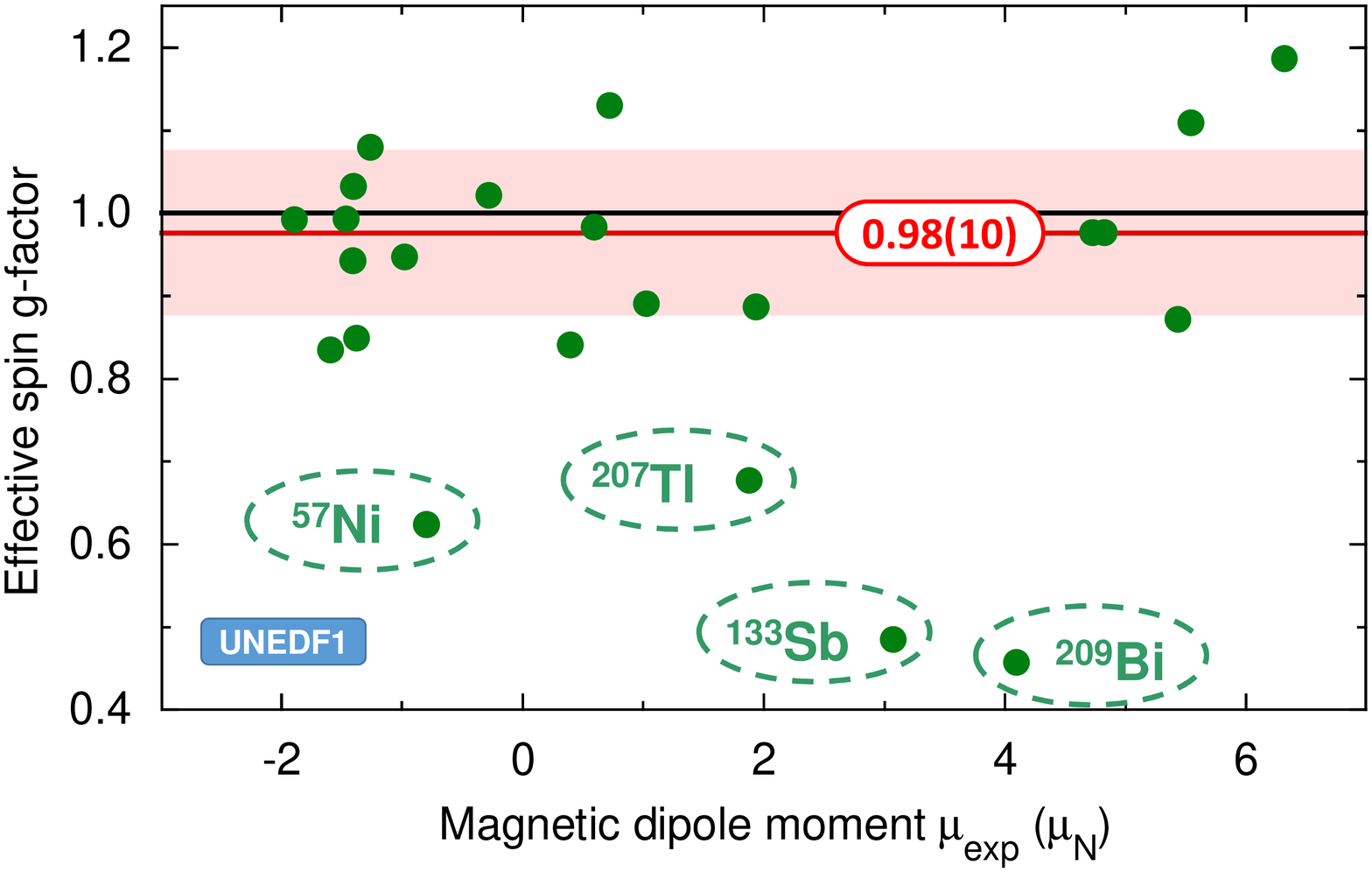}
\end{center}
\caption{Effective spin $g$-factors $g_{\text{eff}}$,
Eq.~(\ref{eq:geff}), that would have been needed for bringing the
calculated UNEDF1 magnetic dipole moments $\mu$ to the 23
experimentally measured values.
}
\label{fig:UNEDF1_geff}
\end{figure}

As one can see in {\Figs}~\ref{fig:UNEDF1_D} and~\ref{fig:UNEDF1_M},
the non-local functionals D1S and N3LO, which have not been
explicitly adjusted to time-odd observables, are characterized by
somewhat lower values of the Landau parameter $g_0'$~\cite{(Idi17)}
and deliver values of the magnetic moments in between of those for
UNEDF1 and Schmidt values, most often away from data. This calls for
performing such adjustments of novel non-local functionals, even
though parameters thereof cannot be explicitly separated into two
distinct classes characterizing the time-even and time-odd mean-field
sectors.

In many theoretical approaches determining nuclear magnetic dipole moments $\mu$ so far,
the bare spin gyromagnetic factors of Eq.~(\ref{eq:miu}) were multiplied
by the so-called effective spin $g$-factor $g_{\text{eff}}$, that is,
\begin{equation}
\label{eq:geff}
g_s^n=-3.826 \,\mu_N\times g_{\text{eff}}, \quad ~g_s^p=+5.586\,\mu_N\times g_{\text{eff}}.
\end{equation}
The DFT results obtained in this work do not
support the necessity of using an effective spin $g$-factor for the
description of nuclear magnetic dipole moments. Indeed, in
{\Fig}~\ref{fig:UNEDF1_geff} we show values of $g_{\text{eff}}$ that
would have been needed for bringing the calculated UNEDF1 magnetic
dipole moments to measured values. One can clearly see that the
average value of $\bar{g}_{\text{eff}}=0.98(10)$, determined without
the outlier cases identified in {\Figs}~\ref{fig:UNEDF1_Q}(b) and~\ref{fig:UNEDF1_Mg0}(b),
is compatible with $g_{\text{eff}}=1$ and within $\pm20\%$ covers all results.

Very small outlier values of $g_{\text{eff}}\leq0.7$ rather call
for a specific configuration-mixing explanation (as already noted
long ago~\cite{(Bli54),(Blo65)} and widely discussed thereafter) than
for a global modification of the single-particle magnetic-moment
operator. We note here that deformed and polarised DFT states
automatically include admixtures of large classes of states that in
the spherical symmetry must be added explicitly. Nevertheless, mixing
of different deformed and polarised DFT
configurations~\cite{(Sat16e)} is an option that still needs to be
considered.

In conclusion, we showed that the nuclear DFT provides a correct
global description of nuclear electric quadrupole and magnetic dipole
moments in one-particle and one-hole neighbours of doubly magic
nuclei. A fair agreement of calculated magnetic dipole moments with
data was achieved by adjusting one coupling constant in the time-odd
mean-field sector of the nuclear functional. An essential element of
the agreement is the self-consistent spin and current polarisation,
operating in a whole single-particle phase space, which induces
correct dipole moments for bare spin gyromagnetic factors. Our study
indicates that the use of effective $g$-factors in previous DFT
approaches~\cite{(Bon15),(Per21),(Bar21)} may be unjustified and
that the studies neglecting the time-odd mean
fields~\cite{(Bor17),(Per21),(Bar21)} may be missing an important
element of the description. The effective $g$-factors, commonly used
in the valence-space approaches, like the shell model, can probably be
attributed to the missing single-particle phase space. For the
quadrupole moments, the analogous role played by the effective
charges has been known since long~\cite{(Boh69a)}; here we
demonstrated it for the magnetic moments.

Our results provide a baseline for
more extensive adjustments in the poorly-known time-odd mean-field sector. Such adjustments may
further reduce the scatter of calculated magnetic dipole
moments around the experimental values. This, in turn, may be the starting point for
the DFT deformed-configuration interaction calculations, especially regarding
large outlier deviations of the magnetic dipole moments currently obtained in $^{57}$Ni,
$^{207}$Tl, $^{133}$Sb, and $^{209}$Bi.

Prospective global DFT adjustments and deformed-configuration interaction would open
up the possibility of quantifying the roles played by the meson-exchange and/or two-body
currents in redefining the magnetic-moment operator.
These are of marked importance for the interpretation of
modern high-precision experiments such as neutrino physics
\cite{(Gys19)} or dark matter searches \cite{Hof15}. Future measurements
of electromagnetic moments of isotopes around doubly magic nuclei
will be important to test our theoretical predictions and further
constrain DFT developments.

\bigskip\noindent
Critical reading of the manuscript by Mike Bentley is gratefully acknowledged.
This work was partially supported by the STFC Grant
Nos.~ST/M006433/1 and~ST/P003885/1, by the Polish National
Science Centre under Contract No.~2018/31/B/ST2/02220, and by
the U.S.\ Department of Energy, Office of Science, Office of Nuclear
Physics under award numbers DE-SC0021176 and DE-SC0021179.
We acknowledge the CSC-IT Center for Science
Ltd., Finland, for the allocation of computational resources.
This project was partly undertaken on the Viking Cluster,
which is a high performance compute facility provided by the
University of York. We are grateful for computational support
from the University of York High Performance Computing
service, Viking and the Research Computing team.

\section*{References}

\bibliographystyle{iopart-num}
\providecommand{\newblock}{}

\end{document}